\documentclass[english,10pt,aps,prd, a4paper,
preprintnumbers, 
floatfix,
nofootinbib,showpacs,superscriptaddress, notitlepage]{revtex4-1} 

\usepackage{graphicx}				
\usepackage{amssymb}
\usepackage{amsmath}
 \pdfoutput=1
\usepackage[usenames,dvipsnames]{color}  
\usepackage{setspace}
\usepackage{braket}
\usepackage{amsmath}
\usepackage{amssymb}
\usepackage[colorlinks=true,citecolor=darkred,urlcolor=darkred, pdfborder={0 0 0}]{hyperref}
\usepackage[normalem]{ulem}
\usepackage{xcolor}
\usepackage{amssymb,wasysym}

\makeatletter
\def\p@subsection{}
\makeatother

\definecolor{darkred}{rgb}{0.6,0,0}

\definecolor{linkcolor}{rgb}{0,0,0.5}

\def\gsim{\raise0.3ex\hbox{$\;>$\kern-0.75em\raise-1.1ex\hbox{$\sim\;$}}}
\def\lsim{\raise0.3ex\hbox{$\;<$\kern-0.75em\raise-1.1ex\hbox{$\sim\;$}}}

\def\beqn#1{\begin{equation}\label{#1}}
\def\eeqn{\end{equation}}

\def\beqa#1{\begin{eqnarray}\label{#1}}
\def\eeqa{\end{eqnarray}}

\def\Z2{$\mathcal{Z_2}$}

\newcommand {\ignore}[1]{}

\def\321{$\mathrm{SU(3) \otimes SU(2) \otimes U(1)}$ }

\newcommand{\MeV}{{\rm MeV }}

\newcommand{\AddrUnina}{Departimento di Fisica E. Pancini, 
Universit\`a di Napoli Federico II \\
Complesso Universitario di Monte Sant'Angelo, Via Cinthia, Napoli (NA), Italy \\ and INFN, Sezione di Napoli.}
 \newcommand{\AddrLNGS}{INFN, Laboratori Nazionali del Gran Sasso,
67100 Assergi, L’Aquila (AQ), Italy.}
 \newcommand{\AddrSMS}{Scuola Superiore Meridionale, Largo San Marcellino 10, I-80138, Naples, Italy.}

\begin{document}
  
\title{\color{BrickRed} An accurate evaluation of electron (anti-)neutrino scattering on nucleons}
\author{Giulia Ricciardi}\email{giulia.ricciardi2@unina.it}
\affiliation{\AddrUnina}
\author{Natascia Vignaroli}\email{natascia.vignaroli@unina.it}
\affiliation{\AddrUnina}
\author{Francesco Vissani}\email{vissani@lngs.infn.it}
\affiliation{\AddrLNGS}\affiliation{\AddrSMS}

 \begin{abstract}
  \vspace{1cm} 

We discuss as accurately as possible the cross section of quasi-elastic scattering of electron (anti-)neutrinos on nucleons, also known as inverse beta decay in the case of antineutrinos.  
We focus on the moderate energy range from a few MeV 
 up to hundreds of MeV, which includes neutrinos from reactors and supernovae. 
We assess the uncertainty on the cross section, which is relevant to experimental advances and increasingly large statistical samples. We estimate the effects of second-class currents, showing that they are small and negligible for current applications.

\end{abstract}
\maketitle


\section{Introduction}

The cross section of the process $ \bar \nu_{\mbox {\tiny e}} + \mbox {p} \to \mbox{e}^+ + \mbox{n} $ (IBD, from inverse beta decay),
which enabled the first direct observation of (anti-)neutrinos \cite{Cowan:1956rrn}, is still essential today at relatively low energies for detectors that are based on water or hydrocarbons, i.e. the most commonly used detectors such as scintillators or Cherenkov light detectors.

The most accurate estimates available today were described about 20 years ago by Beacom and Vogel \cite{Vogel:1999zy} 
and by Strumia and Vissani \cite{Strumia:2003zx}.
In this paper, we aim to update the discussion, in consideration of some general facts: 1)~the importance of the reaction itself, 2)~the experimental progress
of neutrino detectors, 3)~the progress related to the parameters that determine the reaction itself, 
4) the role of second-class currents (SCCs) \cite{Weinberg:1958ut}, previously omitted, but which has been the subject of a recent debate \cite{Ankowski:2016oyj,Giunti:2016vlh,Ivanov:2017ifp}.

Furthermore, there is a specific quantitative point that deserves to be highlighted: it is important to update not only the value, but also the estimate of the uncertainty on the cross section of the IBD process. Indeed, the conservative estimate of the uncertainty at low energies on this IBD cross section, obtained in \cite{Strumia:2003zx} is 0.4\%. This corresponds to one standard deviation on a sample of 60,000 events and therefore it is potentially relevant in cases where the statistical sample is quite large.
There are at least two cases of this type, which concern reactor antineutrinos (with typical energies of 3 MeV and up to 10 MeV) and those from supernovae
(with typical energies of 20 MeV and 
perhaps up to 60-70 MeV). Let us point out that:
\begin{itemize}
\item Reactor neutrino experiment
Daya Bay \cite{DayaBay:2021dqj}
has collected 3.5 million events already.
Similar considerations apply to the other reactor experiments with very high statistics.
But also the future detector JUNO \cite{JUNO:2015zny}
designed to study events from distant reactors,
with a very large fiducial volume, about 20 kiloton of scintillator liquid,
expects to collect 83 events/day which in 6 years will amount to a total of 180000 events.
\item
As for the future galactic supernova, 
Super-Kamiokande (SK) will collect more than 5000 IBD events with a mass of 32.4 kiloton and for
a typical galactic distance of 10kpc (see e.g. \cite{Vissani:2021jxf}), a number that scales with the square of the distance $(10/d)^2$,
and therefore it can be much larger for relatively closer events. \footnote{It is expected that the typical distance to a supernova event in the Milky Way is 
$10\pm 5$ kpc, see e.g.~\cite{Costantini:2005vh}, 
and that the number of events scales with the inverse of the square of the distance 
to the supernova.}
JUNO also expects to see a similar number of events. 
Furthermore, the future Hyper-Kamiokande detector \cite{Hyper-Kamiokande:2018ofw}
will have a mass about 10 times greater than Super-Kamiokande, and
therefore it will collect a similar number of events 
even in the case that
$d=10$ kpc.
\end{itemize}
On these grounds, it will be important to update the assessment of the error on the cross section today and even more so in the future.

In the next section we present explicit expressions for the cross section including all relevant effects. In section~\ref{Analysis} we estimate the uncertainties in the result in detail.  Some illustrative applications of these results are presented in section~\ref{Applications}, while the last section is devoted to the summary. A few technical results, concerning  in particular second class currents,  are confined to the two appendices.

\section{Neutrino Nucleon Cross Section}
In this section, we calculate the interaction cross section between (anti)neutrinos and nucleons. We first define the most general form of the weak charged current 
(Sect.~\ref{se2:1}) 
including all possible form factors and emphasizing the (usually neglected) second class currents. 
Then we give explicit expressions of the tree-level cross sections
including the main radiative corrections (Sect.s ~\ref{se2:3} and \ref{se2:4}). 

\subsection{The Weak Hadronic Current}\label{se2:1}

We are interested in electron neutrinos and antineutrinos interacting with single nucleons. 
%
 %
The first step towards the cross-section analyses is to write an explicit expression for the  nucleon matrix element of the Standard Model (SM) weak charged current. 
When the matrix elements are taken between nucleon states,
which are composites
of  quarks bound by the strong interactions and have finite dimensions, the couplings, now named  form factors, are no longer constant. Moreover, additional terms arise in the hadron current,  provided that the symmetries of the strong interactions and general principles of Lorentz covariance are respected.

\subsubsection{The Six Form Factors}

  One possible formulation of the most general matrix element of the charged weak current between proton and neutron states, of 4-momenta $p_p$ and $p_n$ respectively, is
\begin{align}
\begin{split}
\mathcal{J}_\mu = \,
&\bar{u}_n \bigg(  f_1 \gamma_\mu +  g_1 \gamma_\mu \gamma_5 + i f_2 \sigma_{\mu\nu}\frac{q^\nu}{2M} + g_2  \frac{q_\mu}{M} \gamma_5  + f_3  \frac{q_\mu}{M} +  i g_3 \sigma_{\mu\nu}\frac{q^\nu}{2M} \gamma_5   \bigg) u_p
\end{split}
\end{align}
The normalisation mass scale is $M=(m_n+m_p)/2$. The form factors $f_1$, $f_2$ and  $f_2$  are generally referred to, respectively, as vector, weak magnetism and scalar. 
The terms including them represent the vector part of the current. The terms including $g_1$, $g_2$ and  $g_2$  represent the axial part of the current.
These six dimensionless form factors  are Lorentz invariant, and 
in general depend upon the 
four-momentum transfer squared $t=q^2=-Q^2$, where  $q=p_n-p_p$. 

We have included all the possible independent vector and axial  terms one can build starting from the basis in the four-dimensional Clifford algebra and using the two independent  4-vectors $q$ and $p_p+p_n$. 
We do not force other symmetries as CVC symmetry on the general form  when $m_p \neq m_n$.
We also do not include terms proportional to $\gamma^\mu q_\mu$ since  the matrix element is on shell.

In the  vector current, a term proportional to  $\bar{u}_n (p^\mu_p+p^\mu_n)u_p$ does not appear, since it can be  re-expressed as a linear combination of the first and third term,  by using the Gordon identity. 
The axial part in a  different formulation, including the dependence on $p_p+p_n$, is given in Appendix \ref{A1}.

\subsubsection{G-parity and Second Class Currents}\label{se2:2}

Weinberg has introduced \cite{Weinberg:1958ut} a classification of the hadronic weak currents according to their properties under a transformation called G-parity. G transformation is defined as a product  of the charge conjugation C and of a rotation in isospin space around the second axis:
\begin{equation}
G = C e^{i \pi I_2}
\end{equation}
Under G-parity, the vector and axial vector currents which transform as
\begin{equation}\label{eq:G1}
G V_\mu G^{-1} = V_\mu \, , \quad  G A_\mu G^{-1} = - A_\mu
\end{equation}
are classified as first-class currents. On the contrary, vector and axial vector currents which transform as
\begin{equation}\label{eq:G2}
G V_\mu G^{-1} = - V_\mu \, , \quad  G A_\mu G^{-1} =  A_\mu
\end{equation}
are classified as second-class-currents (SCC).
The vector and axial vector currents of the SM with form factors $f_1$, $f_2$, $g_1$ and $g_2$ transform as in Eq. (\ref{eq:G1}) and are first class currents, the vector and axial vector currents with form factors $f_3$ and $g_3$ 
transform as in Eq. (\ref{eq:G2}) and are second class currents.
We demonstrate these statements in the Appendix \ref{app:Gparity}.

G-parity is a symmetry for strong interactions, broken by mass differences. Second class currents disappear in the limit of exact flavour $SU(3)$ symmetry.  They also disappear by  assuming invariance under isospin rotation (charge symmetry) if time reversal holds (see e.g. \cite{LlewellynSmith:1971uhs}). However, they have been actively searched for (see e.g. \cite{Fatima:2018tzs}) and  for the sake of generality we include them.


\subsection{Cross Section for $\bar{\nu} p \to e^+ n$}\label{se2:3}

We calculate the cross section for the inverse beta decay, namely the process
\begin{equation}
\bar\nu_{\mbox{\tiny e}} (p_\nu) + \mbox{p}  (p_p) \to  \mbox{e}^+(p_e) + \mbox{n}(p_n)
\end{equation}
It is convenient to introduce the definition for the difference and average of neutron and proton masses $m_n$ and $m_p$
\begin{equation}
\Delta = m_n -m_p \approx 1.293 \, \text{MeV}  \qquad M =\frac{m_n+m_p}{2} \approx  938.9 \, \text{MeV}
\end{equation}
This charged-current quasielastic (CCQE) interaction is the dominant reactions of electron antineutrinos till 2 GeV;
for the reasons explained in the introduction, 
we will be especially interested in the low energy part of this region. 
%
The energy threshold is at
$$
E_\nu\ge  E_{\mbox{\tiny thr}} = \frac{(m_n+m_e)^2-m_p^2}{2 m_p}=1.806 \mbox{ MeV}
$$
Neglecting a region of 1 eV immediately above the threshold  \cite{Strumia:2003zx} 
the possible values of the neutrino energies, as a function of the 
positron energies $E_e$, 
can be easily obtained from the following expression, 
$$
E_\nu= \frac{E_e+\delta}{1 - \frac{E_e-p_e \cos\theta}{m_p}} \mbox{ where }\delta=\frac{m_n^2-m_p^2-m_e^2}{2 m_p}=1.294\mbox{ MeV}
$$
simply by varying the positron emission angle 
$\theta$ in the entire range 0 to $\pi$.\\
The differential cross section is given by
\begin{equation}
\frac{d\sigma}{dt} =  \frac{G_F^2 \cos^2\theta_C }{64 \pi (s-m_p^2)^2}   \,  \overline{ |\mathcal{M}^2|}
\end{equation}
where $G_F$ is the Fermi coupling and $\theta_C$ is the Cabibbo angle
(that is linked to the $u-d$ element of the CKM matrix by the equality $ \cos\theta_C=V_{\mbox{\tiny ud}}$). The `core' of the matrix element includes the hadronic current that we have 
just examined:
\begin{align}
\begin{split}
\mathcal{M} = &\, \bar{v}_\nu \gamma^a(1-\gamma_5) v_e \cdot \\
&\bar{u}_n \bigg(  f_1 \gamma_a +  g_1 \gamma_a \gamma_5 + i f_2 \sigma_{ab}\frac{q^b}{2M} + g_2  \frac{q_a}{M} \gamma_5  + f_3  \frac{q_a}{M} +  i g_3 \sigma_{ab}\frac{q^b}{2M} \gamma_5   \bigg) u_p
\end{split}
\end{align}
with $q=p_\nu-p_e=p_n-p_p$. 
Note that $f_3, g_3$ terms are  second class currents.
%
%
A straightforward calculation gives
\begin{equation}
 \overline{ |\mathcal{M}^2|} = A_{\bar \nu}(t) - (s-u)  B_{\bar \nu}(t) + (s-u)^2 C_{\bar \nu}(t) 
\end{equation}
where $s = (p_\nu + p_p)^2$, $t =q^2= (p_\nu - p_e)^2<0$, 
$u = (p_\nu - p_n)^2$ are the usual Mandelstam variables and
\begingroup
\allowdisplaybreaks
\begin{align}\label{eq:ABC}
 A_{\bar \nu} =&\, (t-m^2_e) \bigg[ 8|f^2_1| (4 M^2+t+m^2_e) + 8|g^2_1| (-4 M^2+t+m^2_e) +2|f^2_2| (t^2/ M^2+4t+4m^2_e)  \nonumber \\
& \; + 8m^2_e t  |g^2_2|/M^2 + 16 \text{Re}[f^*_1 f_2](2t+m^2_e) + 32 m^2_e \text{Re}[g^*_1 g_2] \bigg]  \nonumber \\
& - \Delta^2 \bigg[ (8 |f^2_1| + 2 t  |f^2_2|/M^2)(4 M^2+t-m^2_e) + 8|g^2_1| (4 M^2-t+m^2_e) +8m^2_e |g^2_2| (t - m^2_e)/M^2   \nonumber \\
&+  16 \text{Re}[f^*_1 f_2](2t-m^2_e) + 32 m^2_e \text{Re}[g^*_1 g_2]  \bigg] -64 m^2_e M\Delta \text{Re}[g^*_1(f_1+f_2)] + A_{\mbox{\tiny SCC}} \nonumber \\
B_{\bar\nu} =&\, 32 t \text{Re}[g^*_1(f_1+f_2)] + 8 m^2_e \Delta (|f^2_2|+ \text{Re}[f^*_1f_2 +2 g^*_1 g_2])/M + B_{\mbox{\tiny SCC}} \nonumber \\
C_{\bar\nu} =&\, 8 (|f^2_1| + |g^2_1| ) - 2 t |f^2_2|/M^2 +  C_{\mbox{\tiny SCC}} \, ,
\end{align}
\endgroup
where $m_e$ is the electron mass and the neutrino mass has been neglected. The terms $A_{\mbox{\tiny SCC}}, B_{\mbox{\tiny SCC}}, C_{\mbox{\tiny SCC}}$ are generated by second class currents.
When they are zero we recover the expression in \cite{Strumia:2003zx}.
Their expressions are:
\begin{align}
\begin{split}\label{eq:SCabc}
A_{\mbox{\tiny SCC}} = & \,-2 \,  t \, (4-t/M^2)\bigg[4 m^2_e |f^2_3|+|g^2_3|(t - m^2_e)+8\Delta M  \text{Re}[g^*_3 g_1]+ \Delta^2 |g^2_3|\bigg]\\
&+\mathcal{O}(\Delta^3 M)+\mathcal{O}(\Delta \,m^2_e \, t/M)+\mathcal{O}(m^4_e)\\
B_{\mbox{\tiny SCC}} = & \, 8 \, m^2_e \bigg[ 4\,  \text{Re}[f^*_1 f_3] +  \text{Re}[f^*_2 f_3]\, t/M^2 +2 \, \text{Re}[g^*_1 g_3] +\, \text{Re}[g^*_2 g_3]\, t/M^2 + \Delta |g^2_3|/M \bigg] \\
& + 16 \Delta \text{Re}[(f^*_1 +f^*_2) g_3] \, t/M \\
C_{\mbox{\tiny SCC}} = & -2 |g^2_3|\, t/M^2 
\end{split}
\end{align}
The SCC form factor $f_3$  only enters the cross section in terms suppressed by $m^2_e$, or powers of $1/M$, but there are unsuppressed terms involving the axial SCC form factor.

Because of this suppression very precise beta decay measurements have difficulty
limiting the size of $f_3$. Some experimental limits have been set $|f_3|\leq 2$ \cite{Holstein:1984ga}, but higher values up to 4.4 $f_1$ (at zero momentum transfer) are not excluded \cite{Day:2012gb}.
By contrast, the axial second-class current  at zero momentum transfer is reasonably well constrained by studies of beta decay and can be assumed as $g_3/g_1=0.15$ at zero momentum transfer, although in some analyses it can arrive to $g_3/g_1=0.4$ \cite{Day:2012gb}.

\subsection{Cross Section for $\nu n \to e^- p$}\label{se2:4}

For the process 
\begin{equation}
\nu_{\mbox{\tiny e}} (p_\nu) + \mbox{n}  (p_n) \to  \mbox{e}^-(p_e) + \mbox{p}(p_p)
\end{equation}
we obtain: 
\begin{equation}
\frac{d\sigma}{dt}\big(  \nu n \to e^- p \big) =  \frac{G_F^2 \cos^2\theta_C }{64 \pi (s-m_p^2)^2}  \bigg[ A_{\nu}(t) + (s-u)  B_{\nu}(t) + (s-u)^2 C_{\nu}(t) \bigg]
\end{equation}
with:
\begin{align}
\begin{split}
A_{\nu} =  A_{\bar{\nu}} + &16 \frac{\Delta}{M} \bigg(8 \text{Re}[f^*_1 f_3] M^2 m^2_e +2\text{Re}[f^*_1 f_3] m^4_e -2 \text{Re}[f^*_1 f_3] m^2_e t + 4 \Delta \text{Re}[f^*_1 g_3] M m^2_e \\ & +2 \text{Re}[f^*_2 f_3] m^4_e + 4 \Delta \text{Re}[f^*_2 g_3] M m^2_e - 4 \text{Re}[g^*_1 g_3] M^2 m^2_e + 8 \text{Re}[g^*_1 g_3] M^2 t  \\ & + \text{Re}[g^*_1 g_3] m^4_e + \text{Re}[g^*_1 g_3] m^2_e t -2 \text{Re}[g^*_1 g_3] t^2 +2 \text{Re}[g^*_2 g_3] m^4_e \bigg) \\
B_{\nu} = B_{\bar{\nu}} - & 64 \text{Re}[f^*_1 f_3] m^2_e -32 \frac{\Delta}{M}\text{Re}[f^*_1 g_3] t  -16 \frac{m^2_e}{M^2}\text{Re}[f^*_2 f_3] t  -32 \frac{\Delta}{M}\text{Re}[f^*_2 g_3] t \\ & -32 \text{Re}[g^*_1 g_3] m^2_e  -16 \frac{m^2_e}{M^2}\text{Re}[g^*_2 g_3] t  \\
C_{\nu} = C_{\bar{\nu}} \; \, &
\end{split}
\end{align}
where $A_{\bar{\nu}}$, $B_{\bar{\nu}}$, $C_{\bar{\nu}}$ denote the values for the IBD cross section reported in eq.~(\ref{eq:ABC}).

\subsubsection{Radiative Corrections and Final State Interactions}\label{se2:5}

We include the radiative corrections to the cross section \cite{Towner:1998bh, Beacom:2001hr,Kurylov:2001av}
 that are well approximated as \cite{Kurylov:2002vj}
\[
d\sigma(E_\nu,E_e) \to d\sigma(E_\nu,E_e) \left[1 + \frac{\alpha}{\pi} \left(6.00 + \frac{3}{2}\log \frac{m_p}{2 E_e} + 1.2\left( \frac{m_e}{E_e}\right)^{1.5}\right) \right]
\]
where $\alpha$ is the fine-structure constant.
This expression is valid for the range of neutrino energies we are interested in, when the full energy $E_e$ of leptons and bremsstrahlung photons in the final state is measured. Electroweak corrections are already taken into account, once the measured values of the Fermi constant and axial coupling at $Q^2=0$ are used. 
%
For the scattering $\nu_e +n \to p +e$, we include the multiplicative Sommerfeld factor, which accounts for final state interactions
\[
F(E_e)=\frac{\eta}{1-\exp(-\eta)}, \; {\rm with}\;  \eta=\frac{2 \pi \alpha}{\sqrt{1-m^2_e/E^2_e}}
\]

\section{Analysis of the uncertainties}
\label{Analysis}

In this section, we will examine the uncertainties on the cross section. The overall uncertainty for small neutrino energies $E_\nu$, or rather in the limit 
$Q^2\to 0$, depends mainly on a few parameters, and it is discussed in Sect.~\ref{s7:1}.  

The uncertainty at higher energies, on the other hand, depends 
also on the behaviour of the form factors as $Q^2$ varies. 
The traditional treatment of these form factors (see e.g., \cite{Strumia:2003zx}) is based 
on certain global parametrisations, which have been developed for use at energies around $E_\nu\sim\,\mbox{GeV}$, which are much larger than those we are interested in for the detection of reactor or supernova antineutrino signals. 

In the region of energies in which we are interested the following limit applies 
\begin{equation} \label{rangon}
Q^2_{\mbox{\tiny max}}=2 \frac{(E_e+ \delta)(E_e+p_e)}{1- \frac{E_e+p_e}{m_p}}-m_e^2\sim (4 E_e)^2\, .
\end{equation}
Since we expect $E_e\lesssim 50$ MeV in the case of supernova neutrino detection,\footnote{Notice however that in the innermost regions of the supernova, neutrino energies up to three times higher can be reached.} it is reasonable to use 
a simple linear expansion of the form factors. In other words, it is sufficient to parameterise the evolution with $Q^2$ 
by introducing appropriate {\em radii}, one for each form factor, defined according to established conventions and discussed in Sect.~\ref{s7:2}.
We will discuss the effect of the inclusion of second-class currents on the cross-section, showing that, state-of-the-art, their impact is small and unimportant.

The results of the cross-section uncertainty study are summarised in Sect.~\ref{s7:4}.

\subsection{Overall Uncertainty at Low Energies}\label{s7:1}
The main source for uncertainty on the IBD cross section under a few tens of MeVs
is due to the uncertainties of two important constants: the value of the CKM matrix element
 $V _ {\mbox{\tiny ud}}$ and that of the axial coupling $g_1 (0) = \lim_{q^2 \to 0} g_1 (q^2)$.

Since we have independent measurement of these two quantities,
an hyper-conservative estimate of the uncertainty on the cross section is obtained by simply
propagating the effect of the two errors. However, a more complete estimate is also possible, using
the fact that the average life of the neutron depends on the same quantities, and it is precisely measured; in this way, some correlations arise and
must be taken into account.

In this section we discuss the uncertainties of the relevant parameters, and then summarize the procedures for estimating the overall uncertainty. On several occasions it will happen to combine measurement not perfectly consistent with each other. To do this, we will follow the prescription of the PDG \cite{Zyla:2020zbs}:
when the minimum value of $\chi^2$, estimated from the data and the average, exceeds $N-1$, where $ N $ is the number of measurements,
all errors contributing to the result are enlarged by the scale factor
\begin{equation}
S = \sqrt {\frac {\chi^2_{\mbox{\tiny min}}} {N-1}}
\end{equation}
provided that the value of $ S $ is not too large or that there are credible indications of inconsistencies of specific data sets.

\subsubsection {Value of the CKM Matrix Element}

\begin{figure}[th]
\centerline{\includegraphics[width=0.8\textwidth]{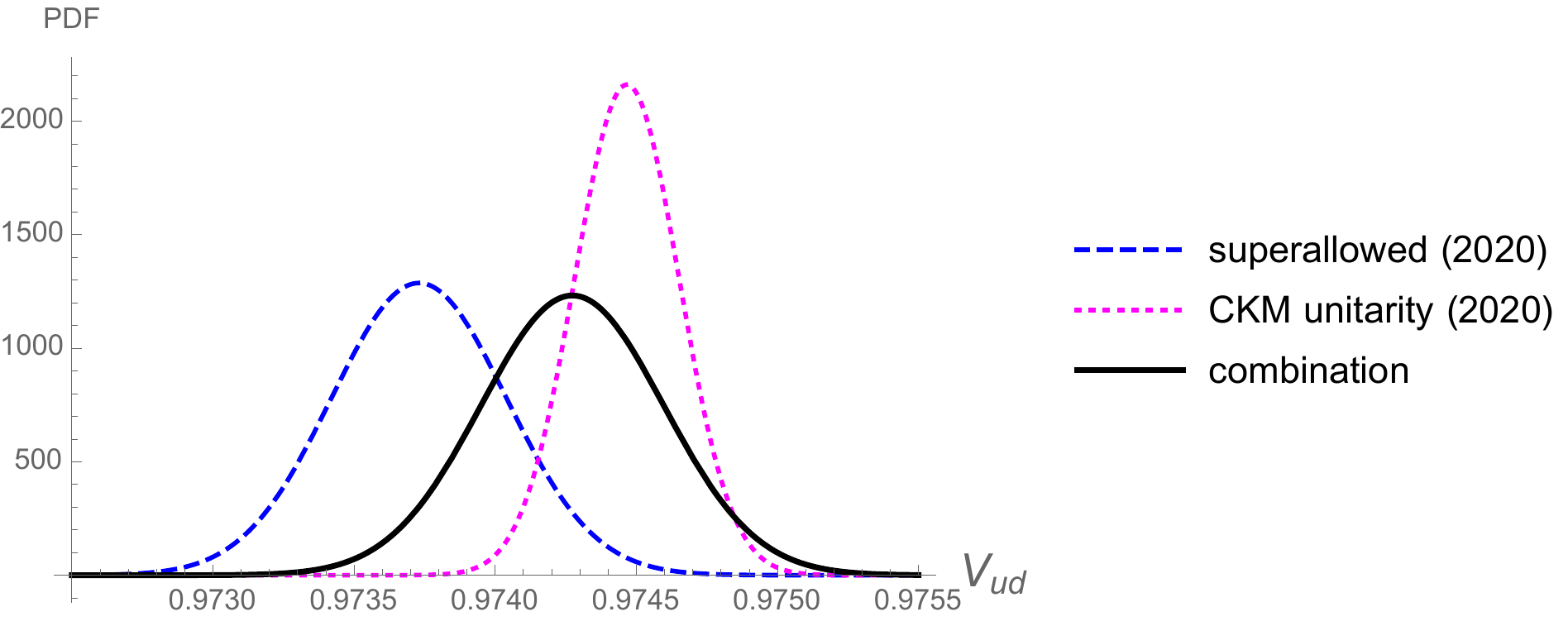}}
\caption{\em Combination of values of $V_{ud}$ extracted from super-allowed decays and unitarity relations.}\label{saunit}
\end{figure}

The most direct way to determine the matrix element we are interested in
takes advantage of the measurements of the super-allowed (s.a.) charged current transitions, which depend only on the better understood vector form factors. From 2006 to 2015, Hardy and Towner produced analyses of all decays this way, including the theoretical factors necessary for the extraction of $V_{\mbox{\tiny ud}} $, see e.g., \cite{Hardy:2014qxa}.
Recently, it has been pointed out that these factors require a more careful treatment
\cite{Seng:2018qru,Czarnecki:2018okw}. This has
led to a critical discussion, which resulted in an higher
conservative  estimate \cite{Hardy:2020qwl}
$|V_{\mbox{\tiny ud}}{\mbox{\small (s.a.)}}| = 0.9737 (3) $ 
which we will adopt later.
A second, more precise, determination follows from noting that,
in the context of the standard model, the CKM matrix is unitary. Therefore, by combining the two measurements of
$|V_{\mbox{\tiny us}}| = 0.2245 (8) $ and
$ |V_{\mbox{\tiny ub}}| = 3.82 (24) \times 10^{- 3} $ reported in
PDG \cite{Zyla:2020zbs}, we get
$V_{\mbox{\tiny ud}}{\mbox{\small (unit)}} = 0.9745 (2) $.
The two previous values are not in perfect agreement, and this has sparked speculation
on the idea that the first value is smaller than the second one  due to a small deviation of the CKM matrix from unitarity. By taking into account
the statistical consistency of this still relatively weak inference, and considering the successes of the standard model at the energies of interest,
we think it is reasonable to assume that this difference merely signals limits to the available interpretations and measurements.
Therefore, we combine the data, obtaining
\begin{equation}
V_{\mbox{\tiny ud}} = 0.9743 (3)  \label{vud:best}
\end{equation}
where the value $ S = 2.0 $ reminds us of the tension between the two measurements. The figure \ref{saunit} shows
the result of the conservative procedure with which we combine the measurements, which, in essence,
prefers the most precise measurement as regards to the central value,
but adopts the error of the less precise one.

\subsubsection {Axial Coupling Measurements}

Using polarized neutrons, see e.g.~\cite{Abele:2008zz},
 it is possible to measure directly
\begin{equation} \lambda = - \frac{g_1 (0)}{f_1 (0)} \, . 
\end{equation}
Eight different measurements are available in the literature \cite{Markisch:2018ndu,UCNA:2017obv,Mund:2012fg,Schumann:2007qe,Mostovoi:2001ye,Liaud:1997vu,Erozolimsky:1997wi,Bopp:1986rt}
and their average is $ \lambda = 1.2755 (5) $ with a scale factor of $ S = 2.3 $.
The latest measurement, published in 2019 by the Perkeo III collaboration \cite{Markisch:2018ndu},
 is much more precise than the others and gives $ \lambda = 1.2764 (6) $.
While the central value and the error of the global mean and of the latter measurement are not very different,
the scale factor $S> 1$ for the mean value indicates a potential problem with the systematics of the previous determinations.
The reference \cite{Czarnecki:2018okw} suggested to exclude the four measurements prior to 2002, namely \cite{Mostovoi:2001ye,Liaud:1997vu,Erozolimsky:1997wi,Bopp:1986rt}
which depended on large and potentially large correction factors
not completely under control. If we adopt this prescription, and also include the result from \cite{Markisch:2018ndu} in the mean,
we get $ \lambda = 1.2762 (5) $ with a scaling factor equal to $ S = 0.7 $; in other words
we have perfect compatibility. On the other hand, rather than   exclude
entirely the measurements prior to 2002, it seems more conservative to include them,
but enlarging their error by a factor 2. In this way we obtain
\begin{equation}
\lambda = 1.2760 (5) \label{lambda:best}
\end{equation}
with a scale factor of $ S = 1.2 $. The value is intermediate between the global average
and the measurement of Perkeo III, and both central values fall within the margin of
one standard deviation.
In light of the current knowledge,
we offer this estimate as a conservative trade-off
and we will use it in the next analysis,
emphasizing that the central value we assumed is in no way crucial to the conclusions that
we will derive.

\subsubsection {The Neutron Lifetime Constraint}

The average lifetime of the neutron depends on the same matrix element of the cross section that interests us
(which is indeed not by chance called ``inverse beta decay") and therefore, we have the 
theoretical prediction \cite{Czarnecki:2019mwq}
\begin{equation}
\frac{1}{\tau_{\mbox{\tiny n}}} = \frac{V_{\mbox{\tiny ud}} ^ 2 \ (1+ 3 \lambda ^ 2)} {4906.4 \pm 1.7 \mbox{s}} \, .
\end{equation}
By propagating the errors, we find the prediction $ \tau_{\mbox{\tiny n}}{\mbox{\small(SM)}} = 878.38 \pm 0.89 $~s.

We can use this relationship together with a measurement of the average life $ \tau_{\mbox{\tiny n}} $ to obtain a constraint on the two quantities of interest.
Now, there are two methods of measurement; in the first, ultra-cold neutrons are trapped and their number is measured over time,
determining the total average lifetime; in the second, the products of the single decay channel predicted by the standard model are observed, using beam neutrons. In other terms,
the beam method measures neutron lifetime by counting the injected neutron and decay product
in the beam.
The results are
$ \tau_{\mbox{\tiny n}}{\mbox{\small (tot)}} = 878.52 \pm 0.46 $~s ($ N = 9 $ and $ S = 1.8 $) and
$ \tau_{\mbox{\tiny n}}{\mbox {\small (beam)}} = 888.0 \pm 2.0 $~s ($ N = 2 $ and $ S = 0.3 $) which clearly disagree.
A priori, it would be possible to hypothesize
an additional neutron decay channel, into particles that are not detected, which would shorten the total average lifetime - a possible way out recently attempted \cite{Strumia:2021ybk}. This would require an agreement between the prediction and
the exclusive measurement, namely that of $ \tau_{\mbox{\tiny n}}{\mbox{\small (beam)}} $,
which is not what is observed: the predicted value $ \tau_{\mbox {\tiny n}} {\mbox {\small (SM)}} $ agrees with 
$ \tau_{\mbox {\tiny n}} {\mbox {\small (tot)}} $, but not with $ \tau_{\mbox{\tiny n}}{\mbox{\small (beam)}} $.

\begin{figure}[t]
\centerline{\includegraphics[width=0.5\textwidth]{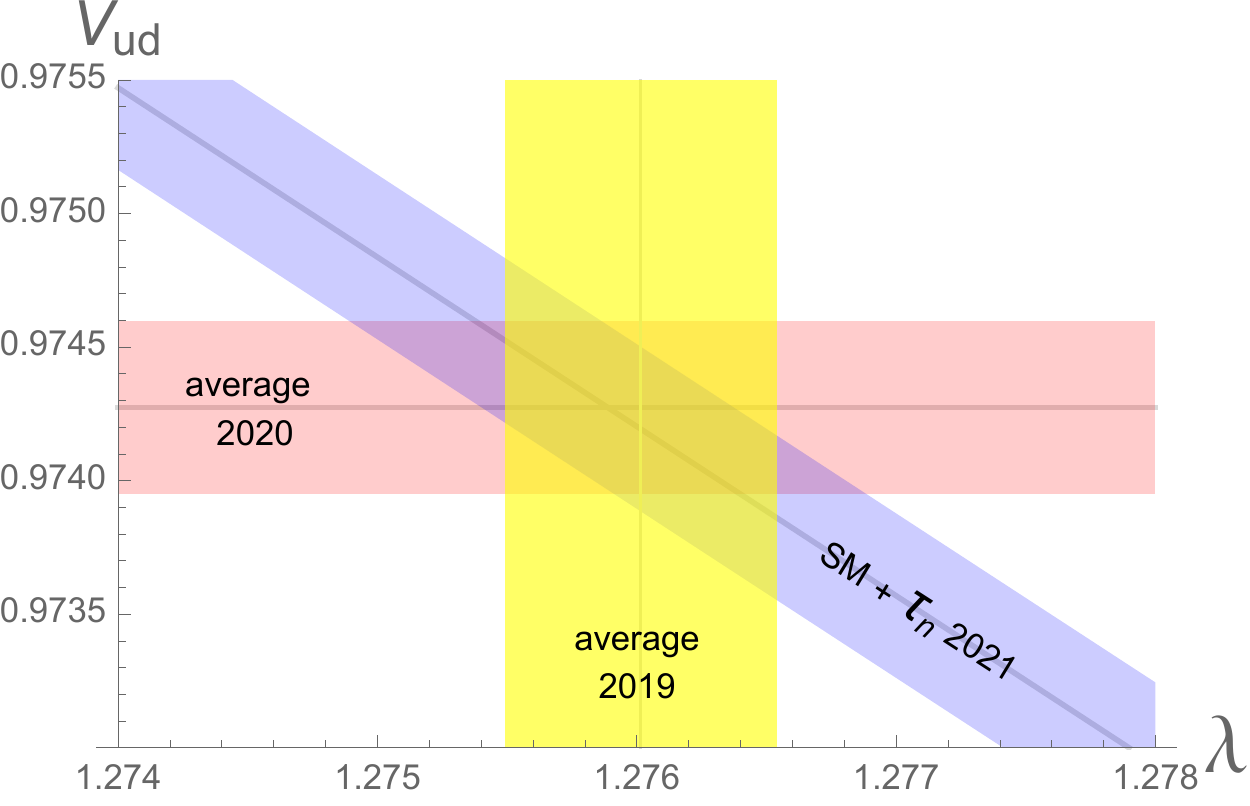}\includegraphics[width=0.5\textwidth]{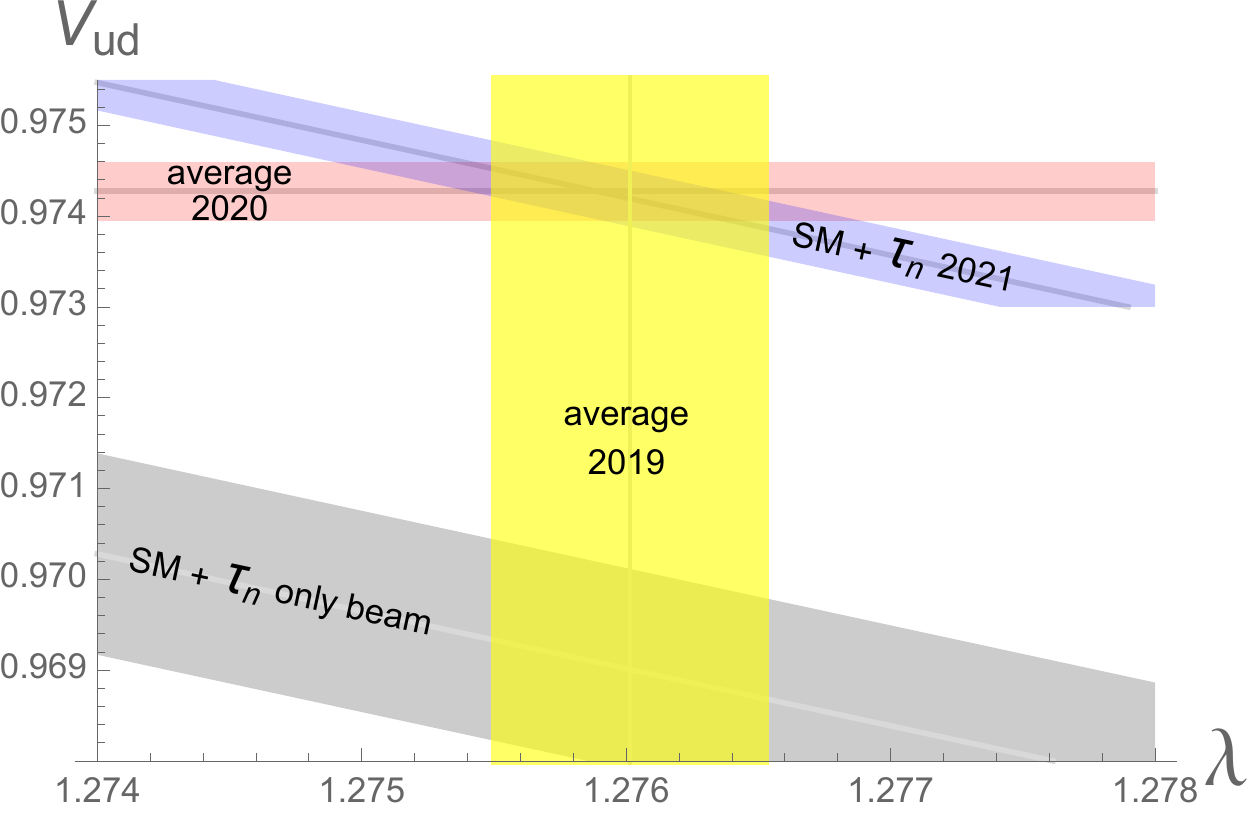}}
\caption{\em Left: illustration of the compatibility,  within the SM,  among the determinations of $\lambda$, $V_{\mbox {\tiny ud}}$ and $\tau_{\mbox {\tiny n}} {\mbox {\small (tot)}}$. Right: 
enlargement of the parameter region to include the prediction of the correlation $ \lambda - V_{\mbox {\tiny ud}}$ (gray band) that follows from the SM assuming the correctness of  measurement $ \tau_{\mbox{\tiny n}}{\mbox{\small (beam)}} $: this is incompatible  with the determinations of  $\lambda$ and $V_{\mbox {\tiny ud}}$.
}\label{fig3}
\end{figure}

For these reasons, we will only use the first set of data,
assuming that the second is affected by a systematic deviation, 
which is not yet fully understood. The observed value of the neutron lifetime 
$ \tau_{\mbox{\tiny n}}{\mbox{\small (tot)}}$, together with
the SM prediction, yield the relationship,
\begin{equation}
  V_{\mbox {\tiny ud}} = \frac{2.36323 (75)} {\sqrt {1+ 3 \lambda ^ 2}} \label{taun:best}
\end{equation}
that we report in  figure \ref{fig3}, 
together with the measurements of $ V_{\mbox {\tiny ud}} $ and of $ \lambda $ mentioned above.
We note that the scale factor relative to the nine measurements included is rather
large, and that there is a trend over time towards decreasing values of $ \tau_{\mbox {\tiny n}} $ \cite{Zyla:2020zbs,Czarnecki:2018okw} specular to that
 of $ \lambda $, which is increasing (in fact,
the most precise measurement, the one obtained in 2021 by UCN $ \tau $ \cite{UCNt:2021pcg}, indicates a value
relatively small,
$ \tau_{\mbox {\tiny n}} {\mbox {\small (UCN)}} = 877.75 \pm 0.28^{+ 0.22}_{- 0.16} $~s,
even if compatible with the average).

\subsubsection{Procedures for Assessing the Uncertainty on the Cross Section}

At this point in the discussion, we can evaluate the uncertainty on the $\sigma $ cross section.
By calculating the derivatives with respect to the parameters of interest, at the point of maximum likelihood,
\begin{equation}
\vec{\xi}=\left( \frac{\partial \sigma}{ \partial V_{\mbox{\tiny ud}}  } \ , \   \frac{\partial \sigma}{ \partial  \lambda  } \right) 
\Big|_{\!\!\begin{array}{c} \mbox{\tiny best}\\[-1.5ex]
\mbox{\tiny fit}
\end{array}}
\end{equation}
we find the uncertainty from the formula
\begin{equation} \label{madaje}
\delta \sigma=   \sqrt{     {\vec{\xi} \, }^t \,   \Sigma^2\,   \vec{\xi}   } 
\quad \mbox{ where }\quad  \Sigma^2=\left( 
\begin{array}{ccc} 
(\delta V_{\mbox{\tiny ud}})^2  & , & \rho\ \delta V_{\mbox{\tiny ud}}\,  \delta \lambda \\[1ex]
 \rho\ \delta V_{\mbox{\tiny ud}}\,  \delta \lambda  & ,  &   (\delta \lambda)^2 
\end{array} \right) 
\end{equation}
where the error matrix $ \Sigma^2 $ is a table of known coefficients (which depends on the data that have been included) and
$ \rho $ is the correlation coefficient. We consider two cases:
\begin{itemize}
\item For an hyper-conservative error estimation procedure, we use
only (\ref{vud:best}) and (\ref{lambda:best})
\begin{equation}\label{eq:cons}
\left\{ 
\begin{array}{l}
V_{\mbox{\tiny ud}} =0.97427(32)\\
\lambda= 1.27601(52)\\
\rho=0 
\end{array}
\right.
\end{equation}
and, as shown in eq.~(\ref{madaje}), we sum in quadrature the effects of the 
two independent errors.
\item
In the case of the full procedure, the one which
includes also the information about the neutron decay, eq.~(\ref{taun:best}),
we have instead
\begin{equation}\label{eq:full}
\left\{ 
\begin{array}{l}
V_{\mbox{\tiny ud}} =0.97425(26)\\
\lambda= 1.27597(42)\\
\rho=-0.53
\end{array}
\right.
\end{equation}
Let us observe that
the central  (best fit) values are almost the same, the errors decrease slightly, and interestingly 
a negative correlation appears, whose effect is to reduce the overall uncertainty on $ \sigma $.
\end{itemize}

\subsubsection{Results}
In the hyper-conservative case, we have:
\begin{equation}
\delta\sigma( {V_{\mbox{\tiny ud}}} ) = 0.66\, \permil \quad
\delta\sigma( \lambda ) = 0.68\, \permil \quad 
\delta \sigma=0.94\, \permil
\end{equation}
In the full treatment case, we have
\begin{equation}
\delta\sigma( {V_{\mbox{\tiny ud}}} ) = 0.53\, \permil \quad
\delta\sigma( \lambda ) = 0.55\, \permil \quad 
\delta \sigma=0.52\, \permil
\end{equation}

In conclusion, we find that: 
\begin{itemize} 
\item 
both errors on $\lambda$ and $V_{\mbox{\tiny ud}}$ are significant and comparable;
\item the correlation has a significant impact, leading to a halving of the error on $\sigma$.
\end{itemize}
The improvement over  2002 estimate \cite{Strumia:2003zx}, when it was estimated that the source of error at low energies amounted to
$ \delta \sigma = 4 \permil$, is noteworthy: 4 times smaller in the hyper-conservative case, almost 8 times in the full case.

\subsection{Effect of Uncertainties on Form Factors}\label{s7:2}

The dependence on $t$ of the form factor can be expressed in several ways. A rather common procedure is to adopt {\em phenomenological} descriptions of the behaviour of the form factors of the nucleons~\cite{LlewellynSmith:1971uhs}. 
Moreover, the form factors  can be constrained  by using analytic methods, crossing symmetry and  global fits, which include several intermediate states and continuum contributions. Finally, the form factors can be calculated {\em ab initio}, in particular, by exploiting  lattice QCD. 

In the present analysis we are mostly concerned with low energy processes. Thus, whatever the expression of the full dependence of the form factor on $t$ is, we  only need the first terms of its Taylor expansion. Furthermore, it should be borne in mind that the phenomenological form factors recalled above - in particular, the dipolar approximation  - are not optimised for the energies we are interested in, and indeed, it has been argued that in some cases of interest the differences are significant. 

With these considerations in mind, the form factors we will use are taken in linearised form.
We follow conventional usage and 
for a generic   form factor $F(t) $  we  
define the corresponding radius $\sqrt{\langle r^2 \rangle}$   
as 
\begin{equation}
\langle r^2 \rangle = \frac{6}{F(0)} \left.  \frac{d F(t)}{dt^2} \right|_{t=0}
\end{equation}
%
%
We recall that the linear expansions are used as a theoretical comparison term for form factors and are usually presented 
in terms of the radii, namely
\begin{equation}
    \frac{ F(Q^2)}{F(0)}   
\equiv 
 1 - \frac{\langle r^2 \rangle \ Q^2}{6}  + \mathcal{O}(Q^4)
\end{equation}
with $Q^2=-t>0$.

\subsubsection{Vector Form Factors $f_1$ and $f_2$}

The linear expansion seems appropriate to assess the uncertainty introduced by the dependence of the form factor on $Q^2$ (or $t$). Indeed, for the energies of interest for supernova neutrino detection, we have $Q^2_{\mbox{\tiny max}}\sim (2 E)^2\lesssim 0.01\mbox{ GeV}^2$
and the higher order terms in $Q^2$ can be safely neglected.
%

The matrix elements of the weak vector currents belong to  the same isotopic multiplet of the matrix element of the iso-vector part of the electromagnetic current. Thus we can estimate the former from the latter by assuming isospin symmetry.

Let us  indicate  Dirac and Pauli electromagnetic  form factors of the proton and of the neutron  with $F_{1,2}^{p,n}(t)$, that 
 are normalized to their electric charge and anomalous magnetic moment when $t=0$: 
$F^p_{1(2)}(0)=1(k_p)$ ad $F^n_{1(2)}(0)=0(k_n)$.
Here $k_p=1.793$ and   $k_n=-1.913$ are the proton and neutron anomalous magnetic moments in units of the nuclear magneton,
and $\xi= k_p-k_n= 3.706$ denotes their difference. 
The isovector parts of the form factors are defined as $F_{1,2}^v=(F_{1,2}^{p}-F_{1,2}^{n})/2$, and 
the corresponding radii, according to \cite{Lin:2021umz}, are 
$\sqrt{\langle r^{v \, 2}_1 }\rangle= 0.751^{+0.002\,+0.002 }_
{-0.001\,  -0.003}\mbox{ fm}$ and $\sqrt{\langle r^{v \, 2}_2} \rangle= 0.880 \pm 0.001 \pm 0.003 \mbox{ fm}$. 
 The nucleon radii accuracy was about 1\% in \cite{Mergell:1995bf}, 
and now it is reduced to about 0.3\% \cite{Lin:2021xrc}. 


%



When we compare these values with the coefficients of $t$ in the dipole approximation,\footnote{
As it is well-known, this is a phenomenological description of  the behaviour of the form factors of the nucleons. 
In this approximation we have
\begin{equation}
    \{f_1, f_2\}=\frac{\{1-(1+\xi)t/4M^2,\xi\}}{(1-t/4M^2)(1-t/M^2_V)^2}\, 
    \label{FFdipoli}
\end{equation}
 thus, $r_1^2= \frac{12}{M_V^2}- \frac{ 3 \xi}{2 M^2}$
and $r_2^2=\frac{12}{M_V^2}+ \frac{ 3 }{2 M^2} $. Note: $M_V^2=  0.71 \, \mbox{GeV}^2$ is a parameter extracted by data.} 
we find $r_1^{\mbox{\tiny dip}}=0.64\, \mbox{fm}$ and $r_2^{\mbox{\tiny dip}}=0.85\, \mbox{fm}$, namely a 
 difference ranging from around 15\% for the former to a 3\% for the latter, which we attribute to the approximated character of the phenomenological description.

Finally, the form factors of the weak charged current are simply given by  $f_{1,2}=2\times F_{1,2}^v$.
Note that conservation of the vector current predicts $f_1(0) = 1$, and isospin-breaking corrections  are expected to play a negligible role. 
By using these central values and  the conversion factor $\hbar c\sim 0.197326... \mbox{GeV}\,\mbox{fm}$,  we  obtain 
$$
f_1\approx 1+ \frac{ (2.41\pm 0.02)\ t}{\mbox{GeV}^2}\qquad
f_2 \approx \xi \left( 1+ \frac{ (3.21\pm 0.02)\ t}{\mbox{GeV}^2} \right)
$$
For supernova neutrino energies, we can estimate  a maximum value of $ E_\nu \simeq E_e \simeq 50$ MeV;  
within these limits, we have at the most 2.4$\permil$ correction on the form factors, which is small.

\subsubsection{The Axial Form Factor $g_1$}\label{Evolution2}
In this section, we quantify the uncertainty in the cross-section due to axial vector form factor $g_1$. As we will see, this form factor  is by far the most important one, being the one that causes the largest uncertainty at high energies. We begin by reviewing its description in the dipole approximation, and then focus on its linear expansion, discussing the value of the so-called axial radius as extracted by a rather conservative approach.


\paragraph{Dipole approximation:}
%
Simple pole  dominance would suggest $g_1 = g_1 (0)/(1 -t/m_{a_1}^2 )$ where the pseudo-vector ${a_1}$ is the pole,  with $J^{PC}=1^{++}$ and  mass 
$m_{a_1}  = (1.23 \pm 0.04) $ GeV, which should give the dominant effect. The usually adopted dipole parametrization  in Eq. \eqref{FFdipoli}, namely
$g_1 = g_1 (0)/(1 -t/M_{A}^2 )^2$, 
provides a better fit of data when $M_{A}$  is left as a free parameter. We expect in this way to parameterize the contribution of other higher states (for which we do not know positions, nor residues). 
Efforts were made in the last decade to extract the value
of the parameter $M_A$. Older data from pion  electroproduction on nucleon experiments, and from neutrino scattering processes off light, intermediate and heavy nuclei have been discussed in \cite{Strumia:2003zx}.
Most of them could be satisfactorily described
with $M_{A} \simeq 1 $ GeV to within a few percent accuracy.
%
%
%
%
\begin{table}
\centering
\begin{tabular}{|cc|}
    \hline
 $M_A$ [\text{GeV}] & \\
   \hline
 1.07(11) &   NOMAD \cite{NOMAD:2009qmu} \\
   1.08(19) &  NOMAD  \cite{NOMAD:2009qmu} 
      \\
      \, $1.19^{+0.09(0.12)}_{-0.10(-0.14)}$ &  MINOS \cite{Dorman:2009zz} \\
       0.99 &   MINER$\nu$A \cite{MINERvA:2013kdn, MINERvA:2013bcy} \\
        1.20(12) &  K2K \cite{K2K:2006odf} \\
       1.36(6) &  MiniBooNE \cite{MiniBooNE:2013qnd} \\
      1.31(3) &    MiniBooNE
     \cite{MiniBooNE:2013qnd}  \\
     $1.26_{-0.18}^{+0.21}$ &  T2K
     \cite{T2K:2014hih}  \\
    \hline
\end{tabular}\caption{ The axial mass $M_A$ as measured in a sample of recent neutrino experiments.}
\label{tab1}
\end{table}
The formal world averaging of $M_A$ values  performed   in 2002 \cite{Bernard:2001rs} from several earlier experiments gives $M_A = 1.026 \pm 0.021$ GeV. 
In 2007, it was updated \cite{Bodek:2007ym}  considering both $\nu_\mu$-Deuterium scattering  and pion electroproduction experiments, resulting in
\begin{equation}
M_A = 1.014 \pm 0.014 \; {\rm GeV} \label{perl}
\end{equation}
The  measurements of the cross-section of the muon neutrino and antineutrino quasi-elastic  scattering processes on a nuclear target (mainly Carbon)  in the (anti)neutrino energy interval 3-100 GeV performed by the NOMAD experiment, running in the years 1995 – 1998, also gave a similar value \cite{NOMAD:2009qmu}.

These values do not include the results from the modern high statistics measurements  performed in the FNAL experiments MiniBooNE (2002-2019), SciBooNE (2007-2008), MINER$\nu$A (2009-2019), MINOS (2004-2016),  and in the (still running) T2K experiments with two near detectors – ND280 (off-axis) and INGRID (on-axis).
They yield larger values, as can be see in Table \ref{tab1}. This has fueled some discussions lately. One possible explanation is that the discrepancy is caused by nuclear effects. Another line of inquiry is based on the
suspicion that the dipole parameterisation may be too restrictive. It  is widely used, especially to model measurements with $Q^2 \gtrsim \mbox{GeV}^2$, but 
 it is purely phenomenological and not necessarily reliable outside where it is probed.
Using the $z$-expansion one generally finds
smaller values and much larger errors for $M_A$, as can be seen in Table \ref{tab2}.
Indeed the dipole fits, being the least flexible, usually tend to give the smallest error, even if they could potentially suffer from a large bias if the true model is in fact not of dipole form.
The dipole is motivated mostly by its simplicity and phenomenological success, whereas the z-expansion is based on a conformal mapping, constructed such that one obtains the largest possible range of convergence for the form factors, treated as
a function of complex arguments.
\begin{table}
\centering
\begin{tabular}{|ccc|}
    \hline
  $z$-expans. &  dipole approx. &   \\
  $M_A$ [\text{GeV}] & $M_A$ [\text{GeV}] & \\
   \hline
 $0.85_{-0.07}^{+0.22} \pm 0.09$ & $1.29\pm 0.05$ & ($\nu_\mu$)  MiniBooNE \cite{Bhattacharya:2011ah} \\
      $0.84_{-0.04}^{+0.12}\pm 0.11$ & $1.27_{-0.04}^{+0.03}$ & ($\bar \nu_\mu$)  MiniBooNE \cite{Bhattacharya:2015mpa} \\
  $0.92_{-0.13}^{+0.12}\pm 0.08$ & $1.00\pm 0.02$ & ($\pi$)  MiniBooNE \cite{Bhattacharya:2011ah} \\
          \hline
\end{tabular}\caption{Comparison between $z$ expansion and dipole approximation in MiniBooNE for QE neutrino/antineutrino-nucleon scattering ($\nu_\mu / \bar \nu_\mu $) and in charged pion electroproduction ($\pi$).}
\label{tab2}
\end{table}

The values of $M_A$ produced by lattice QCD in the last decade, and more frequently in the last years, have generally large errors and  higher values in the dipole ansatz, sometimes finding smaller values using the z-expansion, as shown in Fig 5 of \cite{RQCD:2019jai} by RQCD collaboration, whose own result is $M_A= 1.02(10) $ GeV versus a value in the  dipole approximation of 
$M_A = 1.31(8) $ GeV.

\paragraph{Linear expansion and axial radius:}

In the range of energy we are considering the connection between the z-expansion and the dipole ansatz can be made using the linearized form, as we will detail in the next section.
As already discussed, we can  linearize  the  dependence on $t$  of the  axial form factor $g_1(t)$ in Eq. \eqref{FFdipoli} as we have done in Sect.~\eqref{Analysis} and define a vector axial radius
\begin{equation}
\langle r_{\mbox{\tiny\rm A}}^2 \rangle = \frac{6}{g_1(0)} \left.  \frac{d g_1(t)}{dt^2} \right|_{t=0}
\end{equation}
and, in this framework, we can define  
\begin{equation}
M_{\mbox{\tiny\rm A}}^2 \equiv -2 \frac{{g_1}'(0) }{g_1(0)} = 
\frac{12}{\langle r_{\mbox{\tiny\rm A}}^2 \rangle }
\end{equation}
which can be used to compare the description adopting the traditional  dipole parameterization discussed above with that of any other one, including those better justified theoretically.
For the energies of interest for the detection of supernova neutrinos, we have $Q^2_{\rm max}\sim (2 E)^2\lesssim  0.01\mbox{ GeV}^2$  for  $E<50$ MeV.
Thus, using $M_{\mbox{\tiny\rm A}}^2\sim \mbox{GeV}^2$, 
we find also in this case that  $g_1(Q^2)$ varies by approximately $2$\% and higher order terms in $Q^2$ have a negligible role.





\bigskip

In view of the fact that we are interested in relatively low energies, compared to those for which the dipole parameterisation was developed, it seems important to us to adopt as cautious a procedure as possible. Therefore, following closely the reference \cite{Hill:2017wgb}, we proceed with a critical discussion of the 
main measurements relevant at low energies.
There are three main useful sets of measurements to probe the axial 
radius, namely,  
\begin{itemize}
\item[$\nu\mbox{N}$:] 
direct measurements of charged current interactions, in particular using 
muon neutrinos on Deuterium targets.\footnote{Heavy targets as Carbon are also used to maximize interactions and increase statistics, but 
then nuclear physics corrections are expected much larger.}
Neutral current measurements (using CVC 
and reasonable assumptions on the axial charge) give consistent results \cite{Ahrens:1986xe}. 
The formal errors obtained are  
very small, giving $r_{\mbox{\tiny\rm A}}^2 =0.453\pm 0.023$ fm$^2$ \cite{Hill:2017wgb, Bodek:2007ym},
but the measurements are obtained at higher energies than those that concern us directly, and therefore cannot be used in the case we are interested in 
without an extrapolation. According to \cite{Bhattacharya:2011ah}  - see in particular  Fig. 3 - this extrapolation is not without dangers.  
In this cautious spirit, ref.~\cite{Hill:2017wgb} suggests $r_{\mbox{\tiny\rm A}}^2 =0.46\pm 0.22$ fm$^2$, which does not rely on the dipole approximation
and is consistent with the previous results, but 
has a considerably larger error.


\item[$\mu\mbox{Cap}$:] 
due to crossing invariance, 
the muon capture on proton probes exactly the same form factors at small $Q^2$ and thus 
is extremely interesting for us. The recent MuCap measurement has given results that are quite accurate and consistent with those
of the previous method, with errors of similar magnitude \cite{Hill:2017wgb}. The compatibility of the results 
and credibility of errors allows the combination of results.


\item[$\mbox{e}\to\pi$:] 
Finally, experiments on single pion production by 
electrons on nucleons give extremely precise results, 
and values consistent with previous ones \cite{Hill:2017wgb, Bodek:2007ym}, that - formally - cover precisely the most interesting $Q^2$ region, see e.g.~\cite{Bernard:2001rs}. 
Unfortunately they require using the theory in a regime where its reliability, according to \cite{Hill:2017wgb},  can be doubted. 
\end{itemize}

\paragraph{Summary}

The conclusions of this discussion can be summarized as follows
\begin{equation}\label{eq:rA}
r_{\mbox{\tiny\rm A}}^2=\left\{
\begin{array}{cc}
0.454\pm 0.012 \mbox{ fm}^2\qquad & \mbox{$\nu\mbox{N}$(dipole) \& $\mbox{e}\to\pi$,  \cite{Bodek:2007ym}}\\
0.46\pm 0.16 \mbox{ fm}^2\qquad & \mbox{$\nu\mbox{N}$ \& $\mu\mbox{Cap}$, \cite{Hill:2017wgb}}\\
\end{array}
\right.
\end{equation}
The value of $M_{\mbox{\tiny\rm A}}$ derived from the first (aggressive) procedure coincides with that already discussed in eq.~(\ref{perl}); 
the other one is in perfect agreement as far as the central value is concerned, but 
it is much more cautious when it comes to estimating the  uncertainty range.

In conclusion, despite the many advances, 
 the situation at the moment does not seem entirely different from that described 20 years ago
\cite{Strumia:2003zx}: on the one hand, it is possible to assess the parameter we are interested in 
by adopting a rather aggressive procedure, and in this way the error is minute or negligible, on the other one,  
a prudent, more conservative attitude, is not to be ruled out. 


In the future, theoretical advances (e.g. from lattice QCD) and more precise measurements should be able to significantly 
reduce the error 
even within the conservative approach.\footnote{In  principle, the most straightforward  approach to measuring $r_{\mbox{\tiny\rm A}}^2$ would require to perform scattering experiments with antineutrinos of energies of a few hundred MeV, accessing in this manner $Q^2\sim 0.1-0.3$ GeV$^2$. However, the fact that weak cross sections decrease with energy makes this approach very challenging experimentally.}

\subsubsection{Second-Class Currents and $g_2$}\label{s7:3}

Let us examine the contribution of the second-class currents.
Taking into account the largest possible values for the SCC form factors compatible with data, $f_3=4.4 \,f_1$ and $g_3= 0.4\, g_1$, we evaluate that
the contributions to the total cross sections of electron antineutrinos on protons (inverse beta decay) are at most at the level of 0.3\permil ($E_\nu$/50 MeV), and, as expected, are dominated by $g_3$. 

Summarizing, the state of the art and  the needs of experimental physics seem to well justify the conventional  procedure in which these contributions to the cross section are neglected. In a prudent approach, one could consider including their effect in the estimation of the uncertainty of the cross section; however, the impact of the newly estimated uncertainties is marginal and in practice unimportant. 
These conclusions could be reconsidered only 
if it turned out that the coefficients of the second-class form factors were much larger than those currently estimated.

Similarly, 
we find that the uncertainty associated to $g_2$ is negligible compared to the one from $g_1$. This could be expected, since, as shown in Eq. \eqref{eq:ABC}, $g_2$ enters in the cross section always with a suppression factor $m^2_e$. In particular, if we consider a variation by an order of magnitude in $g_2$, the corresponding modification of the cross section is below the 0.1$\permil$ level at energies $E_\nu \approx$ 50 MeV.

\begin{figure}[t]
\includegraphics[width=0.6\textwidth]{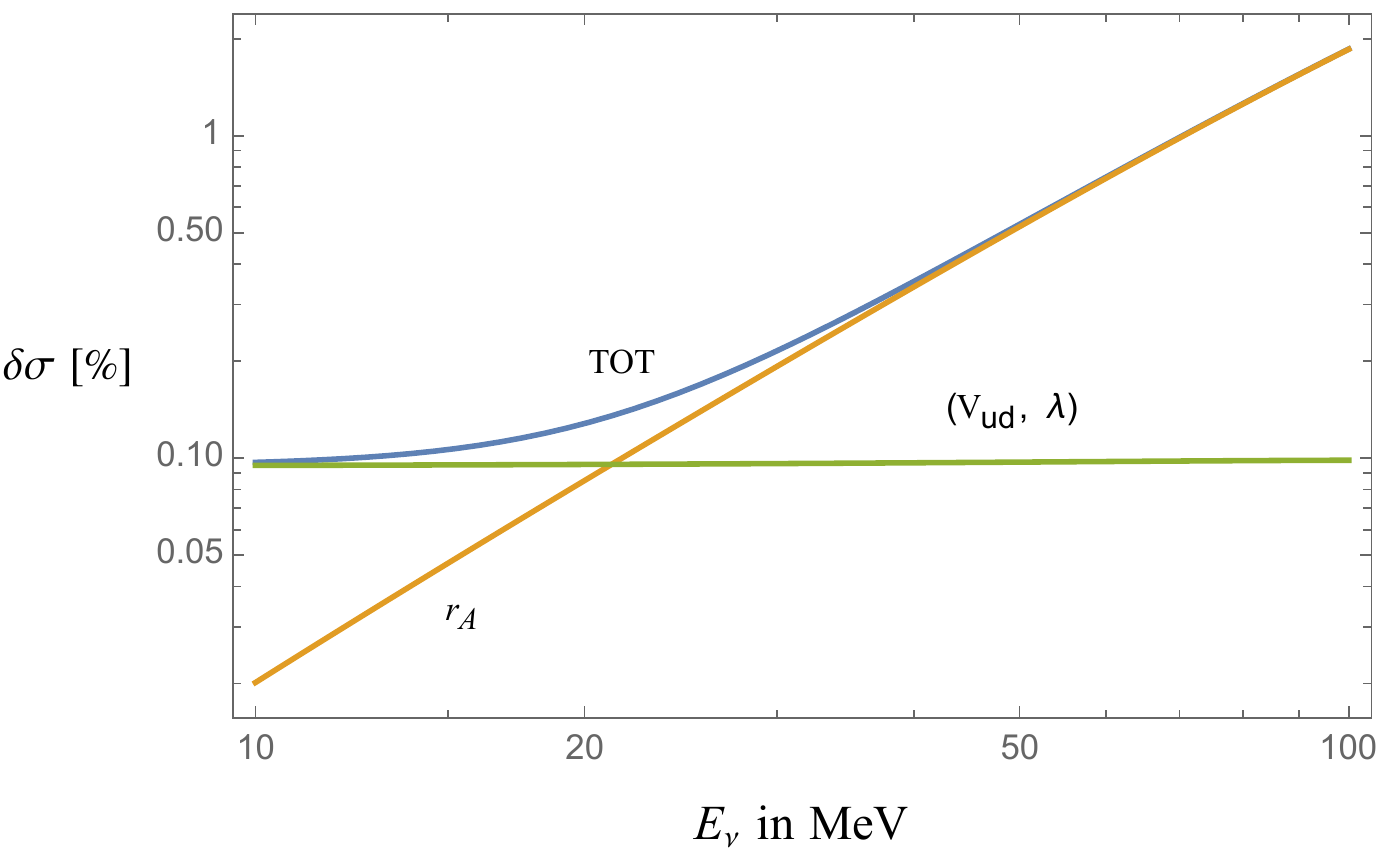}
\caption{\em Overall 1$\sigma$  uncertainty variation of the IBD cross section as a function of the neutrino energy (blue curve). We also indicate the contribution from the errors associated to the $V_{ud}$ and $\lambda$ measurements (green curve), in this case we consider the conservative scenario without correlation induced by the neutron lifetime constraint, and from the uncertainty on $g_1$, encoded in the axial radius (orange curve). Again, we consider the most conservative case. }\label{fig:uncertainty}
\end{figure}

\subsection{Results}\label{s7:4}
Our final results are shown in Fig. \ref{fig:uncertainty}, where we plot the overall 1$\sigma$ uncertainty variation of the IBD cross section as a function of the neutrino energy, adopting the conservative approach.

Note that at higher energies, $E_\nu \gtrsim 20$ MeV, 
the largest source of uncertainty comes from $g_1$, and in particular from its evolution with the energy, which we encoded in the uncertainty on the evaluation of the axial radius $r_A$. If we consider the conservative case in Eq. (\ref{eq:rA}), corresponding to the evaluation of the axial radius from $\nu N$ scattering measurements and $\mu$Cap in the cautious approach of ref.~\cite{Hill:2017wgb}, and a possible range for $r_A$ values within $\pm 1\sigma$ from the central value, we find a maximum shift of 1.8$\permil$  in the total cross section at $E_\nu=20$ MeV, which is of the same order of the total uncertainty on the cross section arising from $V_{ud}$ and $\lambda$, and which increases up to 1.1\% at $E_\nu=50$ MeV. 
However, these evaluations have to be taken as estimates, since we cannot completely rely on the extrapolation from measurements at the GeV scale used in Eq. (\ref{eq:rA}).

As discussed above, the effect of the second class currents and of $g_2$ are 
expected to be negligible.



\section{Applications}\label{Applications}

\begin{figure}[h]
\includegraphics[width=0.55\textwidth]{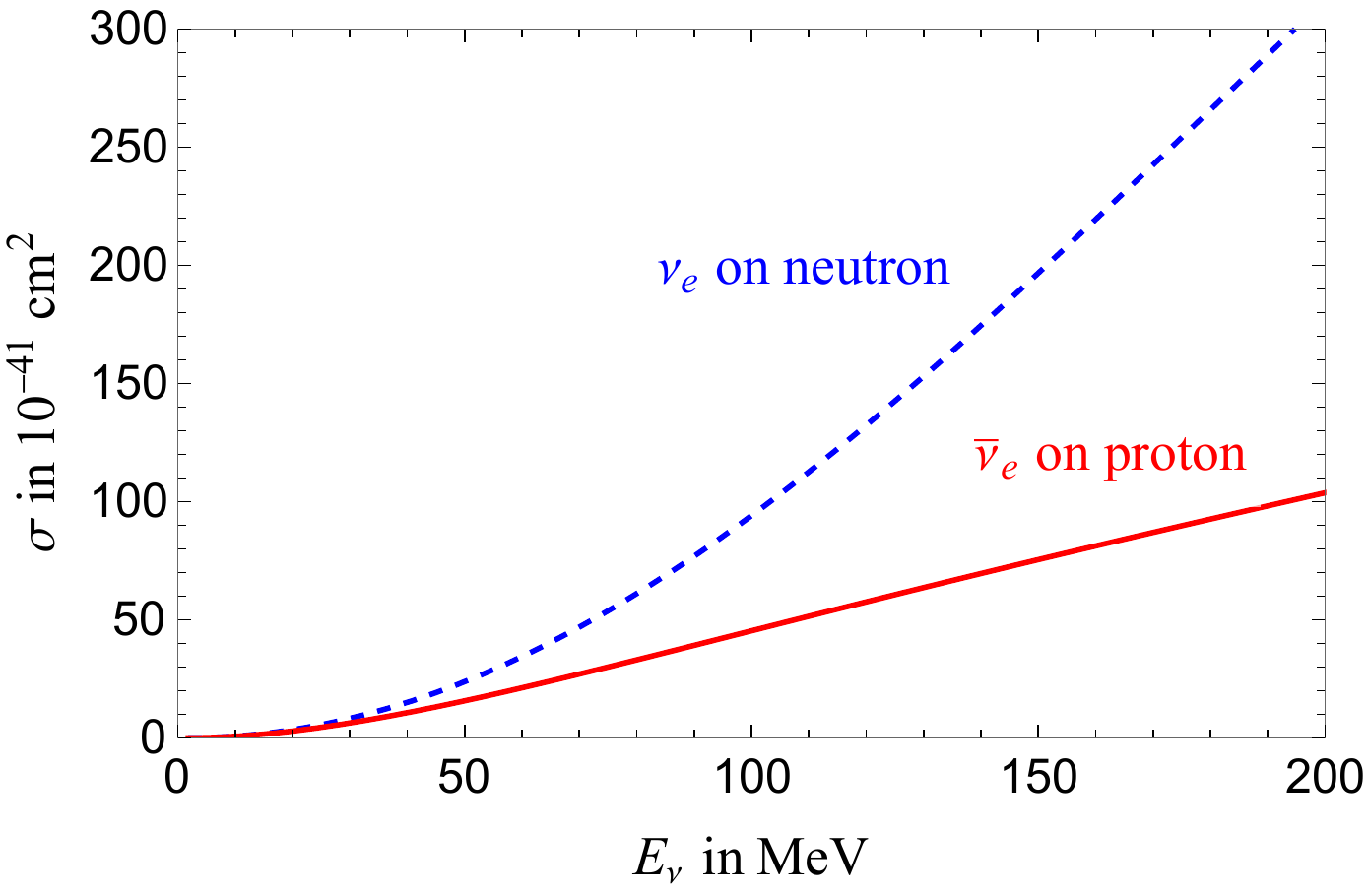}
\caption{\em Total cross sections of quasielastic scatterings at supernova neutrino energies.}\label{fig:xsec-energy}
\end{figure}

\subsection{Estimated Cross Section}

Fig. \ref{fig:xsec-energy} shows the cross sections of quasi-elastic scattering of neutrinos and antineutrinos at supernova neutrino energies, calculated as described above.

The numerical values for the IBD $\bar\nu_e p \to e^+ n$ cross section as function of the neutrino energy are listed in Table \ref{tab:xsecIBD}, and more values in the low energy range [1.9, 12] MeV, of particular interest for reactor experiments, are given in Table \ref{tab:xsecIBD-low}. We consider the hyper-conservative case, corresponding to the central values for the parameters in Eq. (\ref{eq:cons}) and in the second row of Eq. (\ref{eq:rA}).
The cross section reported in \cite{Strumia:2003zx} is in good agreement with ours; at 50 MeV (resp.~100 MeV) 
the value reported in our Tab.~\ref{tab:xsecIBD} is $\sim$0.6\% (resp. $\sim$1\%) higher than that reported in Tab.~1 of that work. The difference is due to the updated parameter values, which were discussed in the previous sections.

\begin{table}
\begin{tabular}{|*{3}{*{2}{|l}|}|}
\hline
$E_\nu$ [MeV] & $\sigma(\bar\nu_e p)$ [10$^{-41}$ cm$^2$] & $E_\nu$ [MeV] & $\sigma(\bar\nu_e p)$ [10$^{-41}$ cm$^2$] & $E_\nu$ [MeV] & $\sigma(\bar\nu_e p)$ [10$^{-41}$ cm$^2$]\\
\hline
$2.$ & $0.00331709$ & $35.$ & $8.42907$ & $68.$ & $25.9144$ \\
$3.$ & $0.0265181$ & $36.$ & $8.86955$ & $69.$ & $26.5082$ \\
$4.$ & $0.0680334$ & $37.$ & $9.31767$ & $70.$ & $27.1041$ \\
$5.$ & $0.127583$ & $38.$ & $9.77319$ & $71.$ & $27.7021$ \\
$6.$ & $0.204738$ & $39.$ & $10.2359$ & $72.$ & $28.302$ \\
$7.$ & $0.299076$ & $40.$ & $10.7055$ & $73.$ & $28.9038$ \\
$8.$ & $0.41018$ & $41.$ & $11.1819$ & $74.$ & $29.5074$ \\
$9.$ & $0.537644$ & $42.$ & $11.6648$ & $75.$ & $30.1126$ \\
$10.$ & $0.681069$ & $43.$ & $12.1541$ & $76.$ & $30.7195$ \\
$11.$ & $0.840063$ & $44.$ & $12.6494$ & $77.$ & $31.3278$ \\
$12.$ & $1.01424$ & $45.$ & $13.1507$ & $78.$ & $31.9375$ \\
$13.$ & $1.20324$ & $46.$ & $13.6577$ & $79.$ & $32.5485$ \\
$14.$ & $1.40667$ & $47.$ & $14.1702$ & $80.$ & $33.1607$ \\
$15.$ & $1.62418$ & $48.$ & $14.688$ & $81.$ & $33.7741$ \\
$16.$ & $1.85542$ & $49.$ & $15.211$ & $82.$ & $34.3886$ \\
$17.$ & $2.10003$ & $50.$ & $15.739$ & $83.$ & $35.004$ \\
$18.$ & $2.35767$ & $51.$ & $16.2718$ & $84.$ & $35.6204$ \\
$19.$ & $2.62801$ & $52.$ & $16.8093$ & $85.$ & $36.2376$ \\
$20.$ & $2.91071$ & $53.$ & $17.3512$ & $86.$ & $36.8556$ \\
$21.$ & $3.20546$ & $54.$ & $17.8974$ & $87.$ & $37.4743$ \\
$22.$ & $3.51193$ & $55.$ & $18.4478$ & $88.$ & $38.0937$ \\
$23.$ & $3.82982$ & $56.$ & $19.0022$ & $89.$ & $38.7136$ \\
$24.$ & $4.15881$ & $57.$ & $19.5604$ & $90.$ & $39.3341$ \\
$25.$ & $4.49861$ & $58.$ & $20.1223$ & $91.$ & $39.955$ \\
$26.$ & $4.84893$ & $59.$ & $20.6877$ & $92.$ & $40.5763$ \\
$27.$ & $5.20946$ & $60.$ & $21.2566$ & $93.$ & $41.1979$ \\
$28.$ & $5.57995$ & $61.$ & $21.8287$ & $94.$ & $41.8199$ \\
$29.$ & $5.96009$ & $62.$ & $22.404$ & $95.$ & $42.4421$ \\
$30.$ & $6.34963$ & $63.$ & $22.9823$ & $96.$ & $43.0645$ \\
$31.$ & $6.7483$ & $64.$ & $23.5635$ & $97.$ & $43.687$ \\
$32.$ & $7.15583$ & $65.$ & $24.1474$ & $98.$ & $44.3096$ \\
$33.$ & $7.57197$ & $66.$ & $24.7339$ & $99.$ & $44.9322$ \\
$34.$ & $7.99647$ & $67.$ & $25.323$ & $100.$ & $45.5549$ \\
\hline
\end{tabular}
\caption{\em Numerical values (in the hyper-conservative case) for the IBD $\bar\nu_e p \to e^+ n$ cross section as function of the neutrino energy.}\label{tab:xsecIBD}
\end{table}

\begin{table}
\begin{tabular}{|*{3}{*{2}{|l}|}|}
\hline
$E_\nu$ [MeV] & $\sigma(\bar\nu_e p)$ [10$^{-41}$ cm$^2$] & $E_\nu$ [MeV] & $\sigma(\bar\nu_e p)$ [10$^{-41}$ cm$^2$] & $E_\nu$ [MeV] & $\sigma(\bar\nu_e p)$ [10$^{-41}$ cm$^2$]\\
\hline
$1.9$ & $0.00190183$ & $5.3$ & $0.1489$ & $8.7$ & $0.497711$ \\
$2.$ & $0.00331709$ & $5.4$ & $0.156356$ & $8.8$ & $0.510862$ \\
$2.1$ & $0.00484225$ & $5.5$ & $0.163987$ & $8.9$ & $0.524173$ \\
$2.2$ & $0.00652675$ & $5.6$ & $0.171791$ & $9.$ & $0.537644$ \\
$2.3$ & $0.00838533$ & $5.7$ & $0.179769$ & $9.1$ & $0.551275$ \\
$2.4$ & $0.0104239$ & $5.8$ & $0.187919$ & $9.2$ & $0.565064$ \\
$2.5$ & $0.0126452$ & $5.9$ & $0.196243$ & $9.3$ & $0.579013$ \\
$2.6$ & $0.0150505$ & $6.$ & $0.204738$ & $9.4$ & $0.59312$ \\
$2.7$ & $0.0176403$ & $6.1$ & $0.213406$ & $9.5$ & $0.607386$ \\
$2.8$ & $0.0204149$ & $6.2$ & $0.222245$ & $9.6$ & $0.621809$ \\
$2.9$ & $0.0233742$ & $6.3$ & $0.231255$ & $9.7$ & $0.636389$ \\
$3.$ & $0.0265181$ & $6.4$ & $0.240435$ & $9.8$ & $0.651126$ \\
$3.1$ & $0.0298462$ & $6.5$ & $0.249786$ & $9.9$ & $0.666019$ \\
$3.2$ & $0.0333584$ & $6.6$ & $0.259306$ & $10.$ & $0.681069$ \\
$3.3$ & $0.0370542$ & $6.7$ & $0.268996$ & $10.1$ & $0.696274$ \\
$3.4$ & $0.0409332$ & $6.8$ & $0.278854$ & $10.2$ & $0.711634$ \\
$3.5$ & $0.0449951$ & $6.9$ & $0.288881$ & $10.3$ & $0.72715$ \\
$3.6$ & $0.0492395$ & $7.$ & $0.299076$ & $10.4$ & $0.74282$ \\
$3.7$ & $0.0536659$ & $7.1$ & $0.309439$ & $10.5$ & $0.758644$ \\
$3.8$ & $0.058274$ & $7.2$ & $0.319969$ & $10.6$ & $0.774622$ \\
$3.9$ & $0.0630633$ & $7.3$ & $0.330665$ & $10.7$ & $0.790753$ \\
$4.$ & $0.0680334$ & $7.4$ & $0.341529$ & $10.8$ & $0.807037$ \\
$4.1$ & $0.0731839$ & $7.5$ & $0.352558$ & $10.9$ & $0.823474$ \\
$4.2$ & $0.0785142$ & $7.6$ & $0.363753$ & $11.$ & $0.840063$ \\
$4.3$ & $0.0840241$ & $7.7$ & $0.375113$ & $11.1$ & $0.856804$ \\
$4.4$ & $0.089713$ & $7.8$ & $0.386638$ & $11.2$ & $0.873697$ \\
$4.5$ & $0.0955806$ & $7.9$ & $0.398327$ & $11.3$ & $0.890741$ \\
$4.6$ & $0.101626$ & $8.$ & $0.41018$ & $11.4$ & $0.907935$ \\
$4.7$ & $0.10785$ & $8.1$ & $0.422197$ & $11.5$ & $0.92528$ \\
$4.8$ & $0.114251$ & $8.2$ & $0.434377$ & $11.6$ & $0.942774$ \\
$4.9$ & $0.120829$ & $8.3$ & $0.44672$ & $11.7$ & $0.960418$ \\
$5.$ & $0.127583$ & $8.4$ & $0.459225$ & $11.8$ & $0.978212$ \\
$5.1$ & $0.134513$ & $8.5$ & $0.471892$ & $11.9$ & $0.996154$ \\
$5.2$ & $0.141619$ & $8.6$ & $0.484721$ & $12.$ & $1.01424$ \\
\hline
\end{tabular}
\caption{\em Numerical values (in the hyper-conservative case) for the IBD $\bar\nu_e p \to e^+ n$ cross section as function of the neutrino energy in the low energy range.}\label{tab:xsecIBD-low}
\end{table}

\subsection{Positron Emission  from IBD Signal in Super-Kamiokande}

We estimate now the energy distribution of positrons generated by supernova anti-neutrino and detected at Super-Kamiokande via the IBD cross section.\\

The anti-neutrino fluence can be described by \cite{Vissani:2014doa}

\begin{equation}\label{eq:flux}
    \frac{dF}{dE_\nu}=\frac{\varepsilon}{4 \pi D^2} \frac{E^2_\nu\, e^{-E_\nu/T}}{6 T^4} \, .
\end{equation}

We consider $\varepsilon=5\times 10^{52}$ erg and $T$=4 MeV.
The positron spectrum is then given by
\begin{equation}\label{eq:positron-rate}
    \frac{dS_e}{dE_e}= N_p \int^{E^{min}_\nu}_{E^{max}_\nu} dE_\nu  \frac{dF}{dE_\nu}(E_\nu) \frac{d\sigma}{dE_e}(E_\nu,E_e) \,\epsilon(E_e) \, ,
\end{equation}
where $N_p$ is the number of protons, which in Super-Kamiokande is particularly large:
\begin{equation}
    N_p = 2 (1-\Upsilon_D)\frac{\pi r^2 h \times \rho_{\rm water}}{m_{\rm H_2 O}}=2.167 \times 10^{33} \, .
\end{equation}
Here $\Upsilon_D=1/6420$ represents the Deuterium contamination, $m_{\rm H_2 O}=2.9915 \times 10^{-23} g$, $h=36.2$ m, $r=16.9$ m, $\rho_{\rm water}=0.998$ g cm$^{-3}$ at 20 °C.

For the efficiency $\epsilon(E_e)$ we use, based on \cite{Vissani:2021jxf},   the same expression, extended to lower energies, adopted for the analysis of SN1987A to describe the response of Kamiokande-II in its entire volume \cite{Vissani:2014doa}:
\begin{align}
    \begin{split}
        & \epsilon(E_e)= \eta(E_e) \frac{1}{2}\left(1+{\rm erf}\left(\frac{E_e-E_{\rm min}}{\sqrt{2} \sigma(E_e)}\right) \right)\\
        & E_{\rm min}=4.5 \, \MeV\\
        & \eta(E_e)=0.93 \left[ 1- \frac{0.2 \MeV}{E_e}-\left(\frac{2.5 \MeV}{E_e}\right)^2\right]\\
        & \sigma(E_e)=1.27 \sqrt{\frac{E_e}{10 \,\MeV}}+\frac{E_e}{10\, \MeV}
    \end{split}
\end{align}

\begin{figure}[t]
\includegraphics[width=0.47\textwidth]{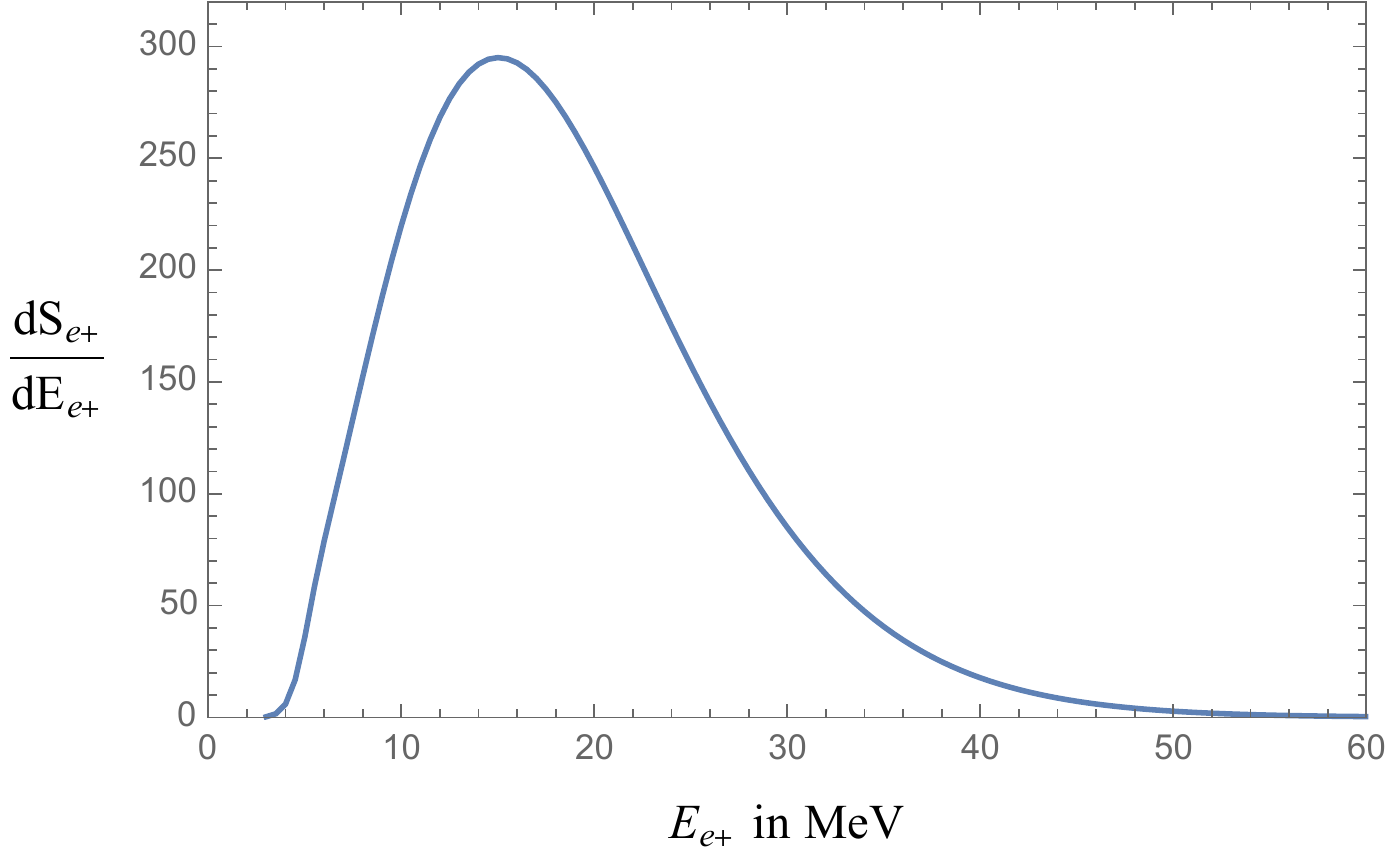}\, \includegraphics[width=0.5\textwidth]{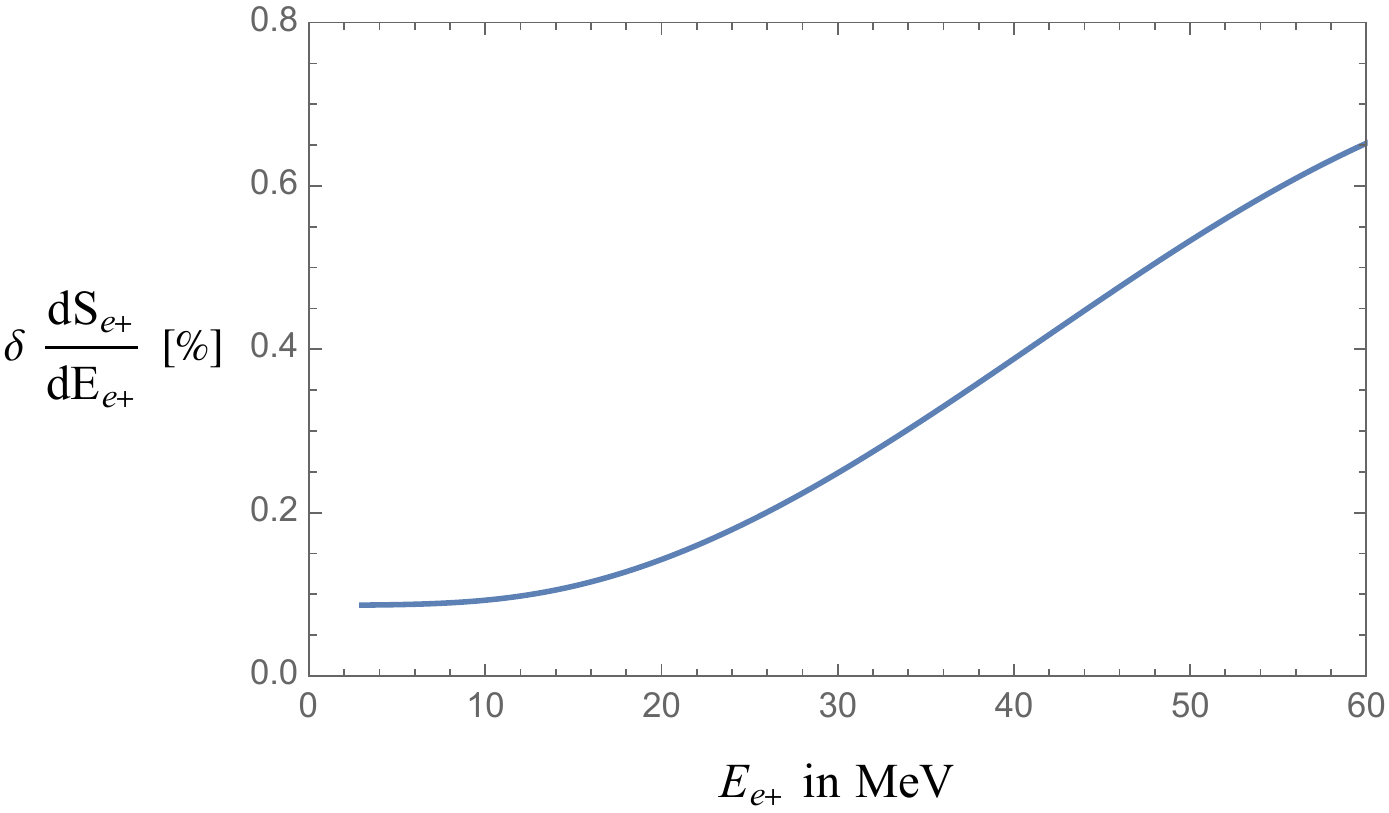}
\caption{\em Energy spectrum of positrons from IBD scattering, considering the antineutrino flux in Eq. (\ref{eq:flux}) (left plot) and the corresponding 1$\sigma$ variation coming from the overall uncertainty on the IBD cross section (right plot).}\label{fig:spectrum}
\end{figure}

The resulting energy spectrum of positrons is plotted in Fig. \ref{fig:spectrum} together with its 1$\sigma$ variation coming from the overall uncertainty on the IBD cross section which we determined as discussed in the previous sections.

\section{Conclusions}

In this paper we have discussed as accurately as possible the cross section of quasi-elastic scattering of electron neutrinos and anti-neutrinos on nucleons.
We focussed on the moderate energy range from a few MeV 
 up to hundreds of MeV, which includes neutrinos from reactors and supernovae. 
We assessed the uncertainty on the cross section, which is relevant to experimental advances and increasingly large statistical samples. We found that
at low energy (relevant for reactor neutrinos) the overall uncertainty on the cross section is dominated by the errors on $V_{ud}$ and $\lambda$. We distinguished two cases: an hyper-conservative approach for the uncertainty estimation and a full procedure, which includes the information on the neutron decay. The uncertainty we find is very small, below per mil. In the first approach, the overall uncertainty on the cross section is 0.94$\permil$, in the second one 0.52$\permil$. Both errors
on $\lambda$ and $V_{\mbox{\tiny ud}}$ are significant and comparable; correlation has a significant impact.\\
At higher energies, 
the largest source of uncertainty comes from the uncertainties in the axial coupling form factor $g_1$. The evaluation of its evolution with energy can be estimated  from measurements at the GeV scale, but this requires an extrapolation to our energies of interest. However, even adopting the cautious approach of ref.~\cite{Hill:2017wgb} (that uses the axial radius obtained from $\nu N$ scattering measurements and $\mu$Cap, but which is less precise) we find that the resultant uncertainty on the cross section is small, about  1.8$\permil$  at $E_\nu=20$ MeV, which is of the order of the total uncertainty on the cross section stemming from $V_{ud}$ and $\lambda$, and which increases up to 1.1\% at $E_\nu=50$ MeV.
We found that the uncertainty associated to $g_2$ is negligible.\\
Finally, we also considered the impact of second-class currents and concluded that their contribution to the cross section can be safely neglected for current needs. 

\section*{Aknowledgments}

We thank Yong-Hui Lin, Eligio Lisi, Laura Marcucci, Ulf-G. Meissner, Gerald A. Miller and Michele Viviani for precious 
discussions.
This work is partially supported by the INFN research initiative ENP.
FV thanks for partial support the Italian Research Grant Number 2017W4HA7S “NAT-NET: Neutrino and Astroparticle Theory Network” under the program PRIN 2017 funded by MIUR.  GR and NV are grateful to the University of Naples  Federico II  and to the Mainz Institute for Theoretical Physics (MITP) of the Cluster of Excellence PRISMA+ (Project ID 39083149), for their hospitality and their partial support during the completion of this work.

\appendix
\section{Alternative Form of the Hadronic Current and Gordon Identities}

We can use an alternative formulation of the hadronic current, as adopted, for example, in \cite{LlewellynSmith:1971uhs}:

\begin{align}
\begin{split}
\mathcal{M} = &\, \bar{v}_\nu \gamma_a(1-\gamma_5) v_e \cdot \\
&\bar{u}_n \bigg(  f_1 \gamma^a +  g_1 \gamma^a \gamma_5 + i f_2 \sigma_{ab}\frac{q^b}{2M} + g_2  \frac{q^a}{M} \gamma_5  + f_3  \frac{q^a}{M} +  g_3  \frac{p_p^a+p_n^a}{M} \gamma_5   \bigg) u_p
\end{split} \label{A1}
\end{align}

Note that, by using a generalization of the Gordon identity, we can express:
\[
\bar{u}_n i  \sigma_{ab}\frac{q^b}{2M} \gamma_5  u_p = \bar{u}_n \bigg( \frac{\Delta}{2 M}\gamma_a \gamma_5 - \frac{p_p^a+p_n^a}{2M} \gamma_5   \bigg) u_p
\] 
This shows that the term $\bar{u} \frac{p_p^a+p_n^a}{M} \gamma_5 u $ used in this alternative formulation does not have definite properties under $G$, unless we restrict to the limit $\Delta=0$, since it can be expressed as a combination of a first class current term and of a second class one.




In this alternative formulation, the terms from second class currents are different from those reported in Eq. \eqref{eq:SCabc} and read:

\begin{align}
\begin{split}
A_{\mbox{\tiny SCC}} = & \,8 \,  t \, (4-t/M^2)\bigg[-m^2_e |f^2_3|+|g^2_3|(m^2_e-t)+2\Delta M  \text{Re}[g^*_3 g_1]+2 \Delta^2 |g^2_3|\bigg]\\
&+\mathcal{O}(\Delta^3 M)+\mathcal{O}(\Delta \,m^2_e \, t/M)+\mathcal{O}(m^4_e)\\
B_{\mbox{\tiny SCC}} = & \, 8 \, m^2_e \bigg[ 4\,  \text{Re}[f^*_1 f_3] +  \text{Re}[f^*_2 f_3]\, t/M^2 - 4 \, \text{Re}[g^*_1 g_3] -  2\, \text{Re}[g^*_2 g_3]\, t/M^2 + 2 \Delta^2/M^2  \, \text{Re}[g^*_2 g_3] \bigg] \\
C_{\mbox{\tiny SCC}} = & -8 |g^2_3|\, t/M^2 + 16 \Delta \text{Re}[g^*_3 g_1]/M + 8 \Delta^2 |g^2_3|/M^2
\end{split}
\end{align}

\section{Properties of the Weak Hadronic Currents Under G-parity}\label{app:Gparity}

We demonstrate the properties of the weak hadronic currents under G-parity for the axial-vector currents (see \cite{Ivanov:2017ifp} for the analogous discussion about the vector currents).

\begin{align}
\begin{split}
& G N G^{-1} = N^{G} = i \tau_2 N^C =  i \tau_2 C \bar{N}^T \\
& G \bar{N} G^{-1} = \bar{N^{G}} = N^T C (- i) \tau_2 
\end{split}
\end{align}

\begin{itemize}

\item
Properties of the current $\bar{N} \gamma_\mu \gamma_5 \tau^{\mp} N$. [$g_1$ term]

\begin{align}
\begin{split}
& G \bar{N} \gamma_\mu \gamma_5 \tau^{\mp} N G^{-1} \\& = N^T C (-i) \tau_2 \gamma_\mu \gamma_5 \tau^{\mp} i \tau_2 C \bar{N}^T = - N^T C \gamma_\mu \gamma_5  \tau^{\pm} C \bar{N}^T \\
&= N^T \gamma^T_\mu \gamma^T_5 \tau^{\pm} \bar{N}^T= -\bar{N} \gamma_\mu \gamma_5 \tau^{\mp} N
\end{split}
\end{align}

This current induces the term with the Lorentz structure $\gamma^\mu \gamma_5$. We shows that it transforms as in Eq. (\ref{eq:G1}) and it is thus a first-class current.

\item
Properties of the current $\partial_\mu (\bar{N} \gamma_5 \tau^{\mp} N)$. [$g_2$ term]

\begin{align}
\begin{split}
& G \partial_\mu (\bar{N} \gamma_5 \tau^{\mp} N) G^{-1} \\& = \partial_\mu (N^T C (-i) \tau_2 \gamma_5 \tau^{\mp} i \tau_2 C \bar{N}^T) = - \partial_\mu (N^T C \gamma_5  \tau^{\pm} C \bar{N}^T) \\
&= \partial_\mu (N^T \gamma^T_5 \tau^{\pm} \bar{N}^T)= -\partial_\mu (\bar{N}\gamma_5 \tau^{\mp} N) 
\end{split}
\end{align}

This current induces the term with the Lorentz structure $q^\mu \gamma_5$. We see that it transforms as in Eq. (\ref{eq:G1}) and it is thus a first-class current.

\item
Properties of the current $\partial_\mu (\bar{N} \sigma_{\mu\nu} \gamma_5 \tau^{\mp} N)$. [$g_3$ term]

\begin{align}
\begin{split}
& G \partial_\mu (\bar{N} \sigma_{\mu\nu} \gamma_5 \tau^{\mp} N) G^{-1} \\& = \partial_\mu (N^T C (-i) \tau_2 \sigma_{\mu\nu} \gamma_5 \tau^{\mp} i \tau_2 C \bar{N}^T) = - \partial_\mu (N^T C \sigma_{\mu\nu} \gamma_5 C \tau^{\pm} \bar{N}^T) \\
&= \partial_\mu (\bar{N} \sigma_{\mu\nu} \gamma_5 (\tau^{\pm})^T N)= \partial_\mu (\bar{N} \sigma_{\mu\nu} \gamma_5 \tau^{\mp} N) 
\end{split}
\end{align}

This current induces the term with the Lorentz structure $i q^\nu \sigma_{\mu\nu} \gamma_5$. We see that it transforms as in Eq. (\ref{eq:G2}) and it is thus a second-class current.

\end{itemize}

\small
\tableofcontents
\end{document}